\documentclass[
aps,
prl,
reprint,
superscriptaddress,
amsmath,
amssymb,
longbibliography,
]{revtex4-1}

\usepackage{graphicx}
\usepackage{dcolumn}
\usepackage{bm}
\usepackage{mhchem}
\usepackage{natbib,hyperref}
\hypersetup{colorlinks=true,allcolors=blue}
\usepackage{ulem}
\usepackage{float}

\renewcommand{\emph}[1]{\textit{#1}}

\begin{document} 
\title{Spin-polarized lasing in a highly photoexcited semiconductor microcavity}
\author{Feng-kuo Hsu}
\thanks{These authors contributed equally to this work.}
\affiliation{Department of Physics and Astronomy, Michigan State University, East Lansing, MI 48824, USA}
\author{Wei Xie}
\thanks{These authors contributed equally to this work.}
\affiliation{Department of Physics and Astronomy, Michigan State University, East Lansing, MI 48824, USA}
\author{Yi-Shan Lee}
\author{Sheng-Di Lin}
\affiliation{Department of Electronics Engineering, National Chiao Tung University, Hsinchu 30010, Taiwan}
\author{Chih-Wei Lai}
\email{cwlai@msu.edu}
\affiliation{Department of Physics and Astronomy, Michigan State University, East Lansing, MI 48824, USA}
\date{\today}
\begin{abstract}
We demonstrate room-temperature spin-polarized ultrafast ($\sim$10 ps) lasing in a highly optically excited GaAs microcavity. This microcavity is embedded with InGaAs multiple quantum wells in which the spin relaxation time is less than 10 ps. The laser radiation remains highly circularly polarized even when excited by \emph{nonresonant} \emph{elliptically} polarized light. The lasing energy is not locked to the bare cavity resonance, and shifts $\sim$10 meV as a function of the photoexcited density. Such spin-polarized lasing is attributed to a spin-dependent stimulated process of correlated electron-hole pairs. These pairs are formed near the Fermi edge in a high-density electron-hole plasma coupled to the cavity light field. 
\end{abstract}
\maketitle 

\section{Introduction}
Circularly polarized light has been used in various applications, ranging from the determination of protein structures with circular dichroism spectroscopy \cite{riehl1986,*bohm1992} and stereoscopic image projection in a polarized three-dimensional (3D) system to spin-dependent electronics (spintronics) and quantum computation \cite{zutic2004,*dyakonov2008,*awschalom2013}. Lasing in semiconductors is generally independent of the spins of electrons and holes, which constitute the gain medium. In semiconductors, spin-polarized electrons (holes) are generated in the conduction (valence) band upon absorption of circularly polarized light above the band-gap, as determined by the optical selection rules \cite{meier1984,*dyakonov1984}. At low photoexcited densities, the photoluminescence (PL) usually comes from spontaneous transitions that do not react back on the electron system. In this case, the degree of circular polarization ($DoCP \equiv \rho_c$) of PL reflects the spin polarization in electrons and holes before radiative recombination \cite{meier1984,*dyakonov1984}. When the spin relaxation rate is fast compared with the radiative recombination, the spin polarization in photoexcited carriers (optical orientation) is typically lost during the absorption--cooling cycle, and unpolarized PL results. Similarly, the $\rho_c$ of the emission in spin-controlled light emitting diodes (spin-LEDs) \cite{fiederling1999,*ohno1999a} is limited by electrical spin injection efficiency. However, in a few spin-controlled lasers, highly circularly polarized radiation can result from partially spin-polarized carriers, provided that radiation from a spin-dependent stimulated process dominates \cite{bresler1987,ando1998,*iba2011,*iba2012,rudolph2003,*rudolph2005,hovel2005,*gerhardt2006,*gerhardt2011,*gerhardt2012,*hopfner2014,holub2005,*holub2007,*holub2007a,*basu2009,*saha2010,blansett2005,adams2009,*schires2012,gothgen2008,*zutic2011,*lee2012,*lee2014,frougier2013,chen2014}.  

In this study, we demonstrate room-temperature spin-polarized ultrafast pulsed lasing in a highly photoexcited planar semiconductor microcavity. The spin-polarized lasing is attributed to a spin-dependent stimulated process of correlated electron-hole pairs formed near the Fermi edge in a high-density electron-hole plasma coupled to the cavity light. The spin-polarized laser studied here has a structure similar to vertical-cavity surface-emitting lasers (VCSELs) \cite{iga1988,*michalzik2013} and microcavities used for studies of exciton-polariton condensates \cite{deng2002,kasprzak2006,balili2007,lai2007,*utsunomiya2008,lagoudakis2008,*lagoudakis2009,*sanvitto2010,*nardin2011,*roumpos2011,bajoni2007,*bajoni2008,*assmann2011,*kammann2012,christmann2012a,schneider2013,*bhattacharya2013,deng2010,*keeling2011,*deveaud-pledran2012,*carusotto2013,*bloch2013,*byrnes2014}. In VCSELs, the lasing energy is typically determined by the bare cavity resonance and has limited energy shifts \cite{chang-hasnain2000,*koyama2006} and linewidth broadening \cite{henry1982,*henry1986,*arakawa1985} with increasing carrier density. The polarization properties of VCSELs are typically affected by crystalline anisotropies \cite{panajotov2013,*ostermann2013}, except for a few spin-controlled lasers \cite{ando1998,*iba2011,rudolph2003,*rudolph2005,hovel2005,*gerhardt2006,*gerhardt2011,*gerhardt2012,*hopfner2014,holub2005,*holub2007,*holub2007a,*basu2009,*saha2010,blansett2005,adams2009,*schires2012,gothgen2008,*zutic2011,*lee2012,*lee2014,frougier2013,chen2014}. On the other hand, exciton-polariton condensates are typically studied in a low photoexcited density regime below the Mott transition \cite{manzke2012,*kappel2005,*klingshirn2012ch20,*klingshirn2012ch21,*klingshirn2012ch22} at cryogenic temperatures. Exciton-polariton condensates generally display considerable spectral blueshifts and broadening with increasing photoexcited density as a result of polariton-polariton interactions, as well as decoherence and fluctuations induced by interactions \cite{wouters2008,*schwendimann2008,*haug2010,*haug2012}. Furthermore, exciton--polariton condensates exhibit diverse polarization properties and spin-dependent phenomena depending on the excitation conditions and materials involved \cite{martin2002,deng2003,*shelykh2004,*cao2008,ballarini2007,baumberg2008,*ohadi2012,vina1996,*vina1999,*ciuti1998,*de-leon2003,*vladimirova2010,*lecomte2014,*sich2014,*takemura2014,kammann2012a,*sekretenko2013,cerna2013,*manni2013,*wouters2013,fischer2014,kavokin2004,*shelykh2006,*shelykh2010,kavokin2012,*kavokin2013,*rubo2013,*flayac2013}. 

\begin{figure}[htb!]
\includegraphics[width=0.48 \textwidth]{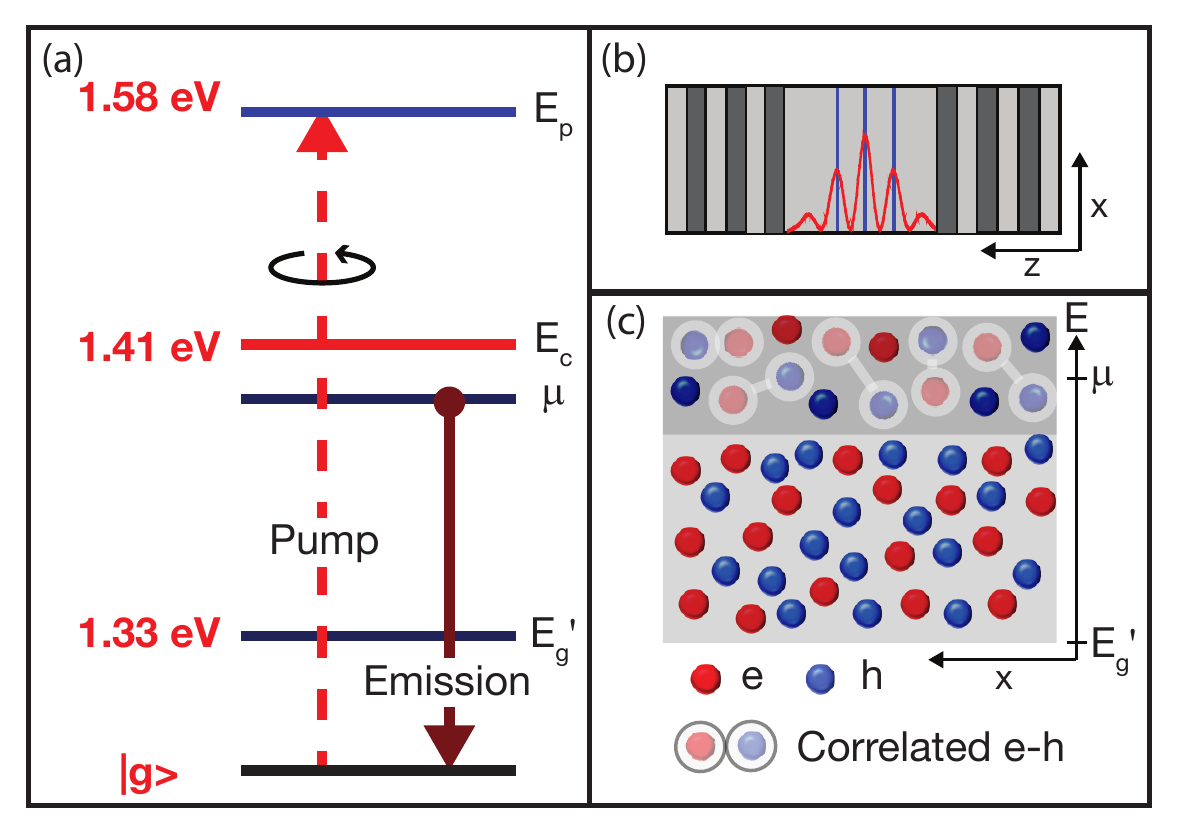}
\caption{\label{fig:energy_diagram} (a) Laser energy diagram. (b) Schematic of the microcavity structure. The sinusoidal red curves represent the amplitude of the cavity light field. The $z$-axis is the growth direction and is parallel to the wavevector of the pump laser. (c) Pictorial representation of the high-density \emph{e--h} plasma in the quantum wells embedded in the microcavity. Correlated \emph{e--h} pairs are formed near the Fermi edge of the \emph{e--h} plasma as a result of effective coupling to the cavity light field.}
\end{figure}

In contrast to conventional spin-controlled lasers (spin lasers) \cite{ando1998,*iba2011,*iba2012,rudolph2003,*rudolph2005,hovel2005,*gerhardt2006,*gerhardt2011,*gerhardt2012,*hopfner2014,holub2005,*holub2007,*holub2007a,*basu2009,*saha2010,blansett2005,adams2009,*schires2012,gothgen2008,*zutic2011,*lee2012,*lee2014,chen2014}, the spin-polarized lasing presented in this study displays (1) substantial \emph{energy blueshifts} of more than 10 meV with increasing photoexcited density, (2) spin-dependent energy splittings in the absence of an external magnetic field, (3) ultrafast \emph{sub-10-ps} pulsed lasing, and (4) a high external quantum efficiency of $\sim10\%$, which matches the fraction of carriers photoexcited in the MQWs. In particular, the lasing energy is largely determined by the chemical potential $\mu$ of the \emph{e--h} plasma coupled to the cavity light field: the lasing energy is not locked to the \emph{bare} cavity resonance. 

\begin{figure}[htb!]
\includegraphics[width=0.4\textwidth]{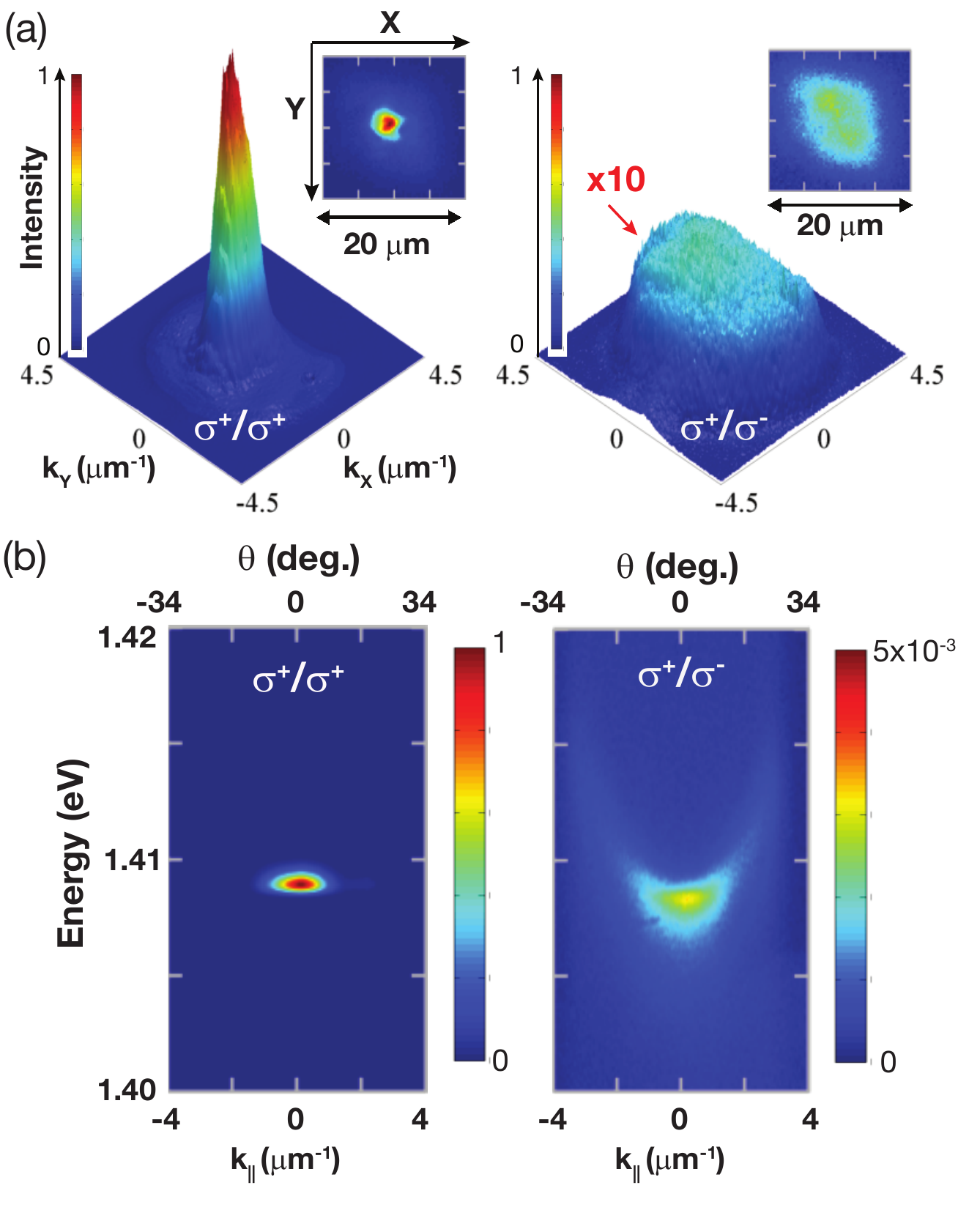}
\caption{\label{fig:kr_images}\textbf{Spin-polarized lasing at room temperature.} (a) Angle-resolved [k-space $(k_X,k_Y)$] luminescence images at the lasing threshold ($P$ = $P_{th}$) for co-circular ($\sigma^+/\sigma^+$, left panel) and cross-circular ($\sigma^+/\sigma^-$, right panel) components. Here, $\sigma^\pm/\sigma^\pm$ represents the polarization of pump/luminescence, respectively. $P_{th}\,\approx$  $2.5\times10^8$ photons \emph{per pulse} (over an area of 80 $\mu$m$^2$), resulting in a photoexcited density $n_{th}\approx 3 \times 10^{12}$ cm$^{-2}$ per QW \emph{per pulse} for an estimated absorption of 10\% for nine QWs. Insets are the corresponding real space (r-space) luminescence images. (b) Energy ($E$) vs. in-plane momentum ($k_\parallel$) dispersions along the $k_Y$ axis ($k_X=0, k_Y=k_\parallel$). 
}
\end{figure}

The room-temperature \emph{e--h} system explored in this work consists mainly of free carriers and high-density \emph{e--h} plasmas as a result of thermal ionization \cite{chemla1984,*chemla1985,schmitt-rink1985a,knox1985,*colocci1990a}. Nevertheless, the presence of a cavity ensures the emergence of correlated \emph{e--h} pairs near the Fermi edge (Fig.~\ref{fig:energy_diagram}) and consequent spin-dependent stimulation that result in ultrafast pulsed lasing with high quantum efficiency. The transient chemical potential $\mu$ can reach the second quantized energy levels in QW ($e2hh2$ transition, $E_g'' \approx$ 1.41 eV), which is more than 80 meV above the QW bandgap ($e1hh1$ transition, $E_g' \approx$ 1.33 eV) and approaches the cavity resonance ($E_c \sim$1.41 eV). At a low photoexcited density, $\mu$ is far off-resonant with respect to $E_c$ (Appendix Fig.~\ref{fig:sample}b--c), and the radiative recombination of \emph{e--h} carriers is largely suppressed. $\mu$ advances toward $E_c$ with increasing photoexcited density, so nonlinearly enhanced luminescence efficiency and lasing eventually result. Moreover, the lasing polarization can be effectively controlled by optical pumping because of a negligible TE-TM mode energy splitting at $k_\parallel$ = 0 (Appendix Fig.~\ref{fig:sample}a). Spin-polarized carriers are optically injected by \emph{nonresonant} ps circularly pump pulses. Below the threshold, luminescence is unpolarized because of sub-10 ps electron-spin ($\tau_s^e$) and hole-spin ($\tau_s^h$) relaxation times \cite{damen1991,*sham1993,tackeuchi1997,amand1994,*hilton2002} that are short compared with the carrier lifetime $\tau_n \gtrsim 1$ ns. Above a critical photoexcited density, laser action commences with a high degree of circular polarization, close to unity ($\rho_c>$0.98) (Fig. \ref{fig:kr_images}). When excited by a \emph{nonresonant elliptically} polarized pump, the $\rho_c$ of the circularly polarized laser radiation can even exceed that of the pump in the limited photoexcited density regime. 

\begin{figure}[htb!]
\includegraphics[width=0.38\textwidth]{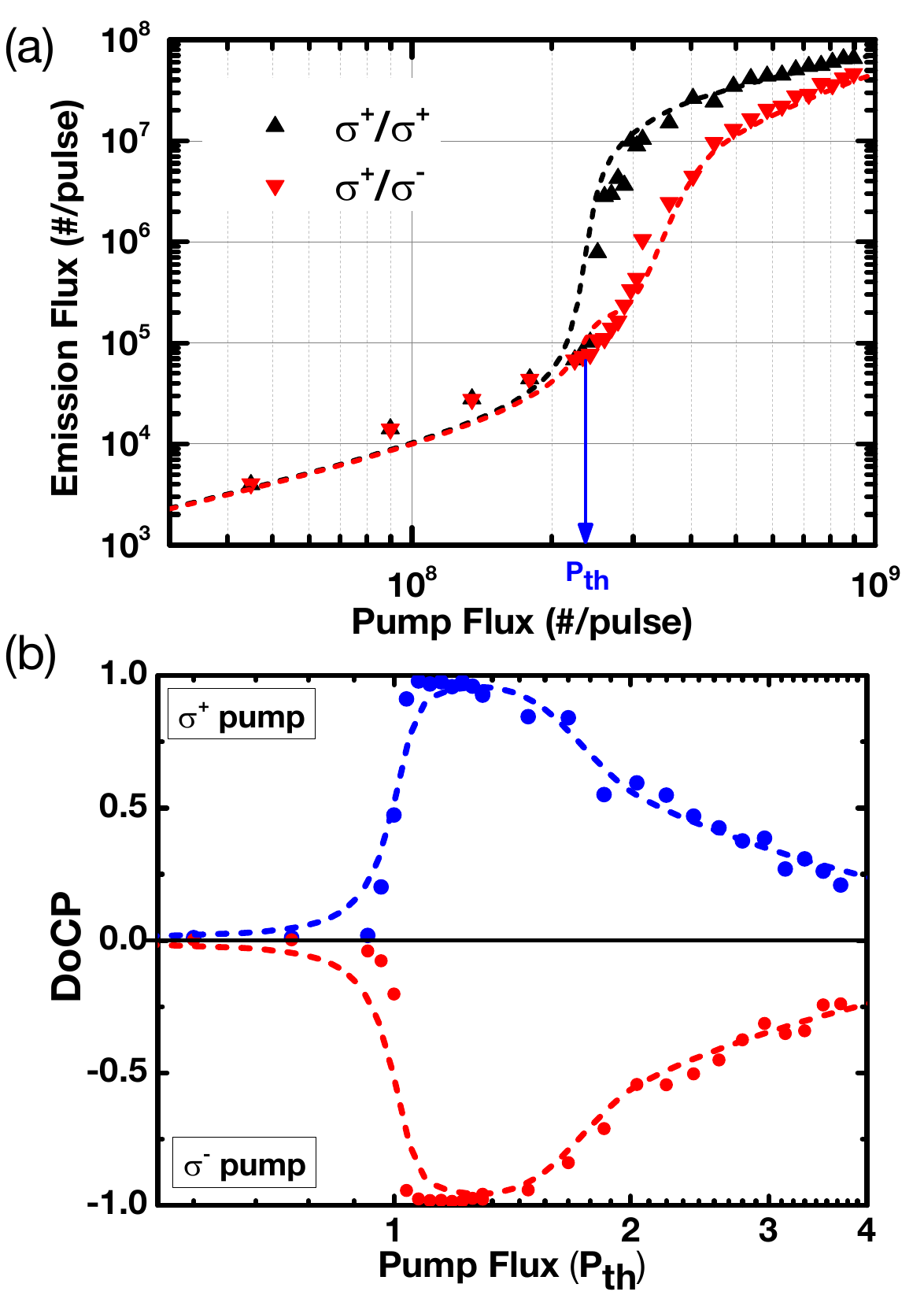}
\caption{\label{fig:lasing}\textbf{Photoexcited density dependence.} (a) Emission flux integrated over $|k_\parallel| < 3$ $\mu$m$^{-1}$ under a  circularly polarized $\sigma^+$ pump. (b) The degree of circular polarization $DoCP$ ($\bar{\rho}_c$), determined from the luminescence integrated near $k_\parallel \approx 0$ ($|k_\parallel|<0.3$ $\mu$m$^{-1}$) under a  $\sigma^+$ or $\sigma^-$ circularly polarized pump. The dashed lines are the calculated emission flux and the $\bar{\rho}_c$, with a spin-dependent stimulated process assumed (see Appendix: Rate-equation Model).} 
\end{figure} 

\section{Results}
The $\lambda$ GaAs/AlAs distributed Bragg reflector (DBR) microcavity examined in this study has three stacks of three InGaAs/GaAs MQWs each, embedded at the antinodes of the cavity light field. We optically pump the sample nonresonantly by using 2 ps Ti:Sapphire laser pulses at $E_p$ = 1.579 eV ($\lambda_p = 785$ nm), which is near a reflectivity minimum (reflectance $\approx$40\%) of the microcavity and about 170 meV above the lasing energy. To investigate laser action in a highly optically pumped microcavity, we vary the laser pump flux by two orders of magnitude and create a photoexcited density ranging from approximately $5\times10^{11}$ to $10^{13}$ cm$^{-2}$ per QW \emph{per pulse}, corresponding to a 2D density parameter $r_s = 1/ (a_0 \sqrt{\pi n_{th}}) \approx$ 5.3--1.2 for $a_0\approx$15 nm in InGaAs QWs \cite{atanasov1994}. To control carrier heating and diffusion, we temporally modulate pump intensity and spatially shape the pump laser beam into a flat-top, respectively (see also the Appendix~\ref{sec:methods}: Methods).

\begin{figure*}[htb!]
\centering
\includegraphics[width=0.85 \textwidth]{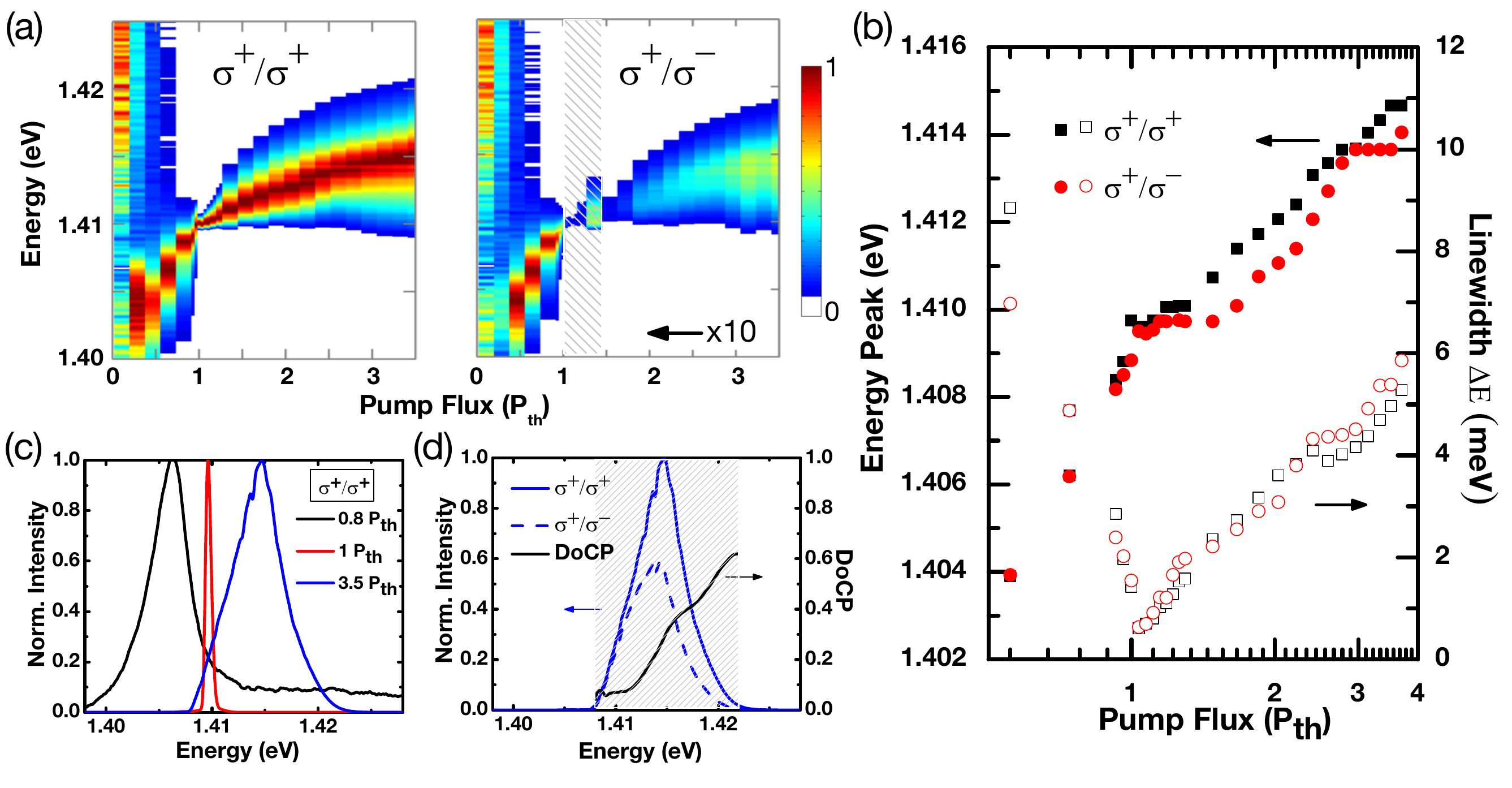}
\caption{\label{fig:spec}\textbf{Time-integrated polarized spectra.} (a) 2D false-color images of microcavity luminescence/lasing spectra vs. the pump flux for co-circular ($\sigma^+/\sigma^+$, left panel) and cross-circular ($\sigma^+/\sigma^-$, right panel) components. Spectra are normalized with respect to the co-circular component ($\sigma^+/\sigma^+$) for each pump flux. Note that the intensities of the $\sigma^+/\sigma^-$ spectra for 0.8 $P_{th}$ $<$ $P$ $<$ 1.3 $P_{th}$ (shaded area) are scaled up by a factor of 10. (b) Spectral linewidths $\Delta$E (FWHM) and peak energies determined from the emission spectra in (a). (c) Selected co-circular time-integrated polarized spectra for $P =$ 0.8 (black), 1.0 (red) and 3.3 (blue) $P_{th}$. Spectral blueshifts and linewidth broadening with the increasing pump flux can be readily identified in these selected spectra. (d) Polarized time-integrated spectra (blue solid line: co-circular; blue dashed line: cross-circular) and spectral $DoCP$ [$\bar{\rho}(E)$, black line] for $P =$ 3.5 $P_{th}$. The time-integrated spectra reveal an apparent spin-dependent energy splitting of 1--2 meV.}
\end{figure*}

First, we characterize the laser in terms of the angular distribution and energy as functions of the in-plane momentum [angle-resolved (k-space) images and $E$ vs. $k_\parallel$ dispersions] (Fig. \ref{fig:kr_images}). At the threshold, the emission of the microcavity investigated in this study becomes angularly and spectrally narrow for the co-circular $\sigma^+/\sigma^+$ component, where $\sigma^+/\sigma^+$ is the helicity of the pump/emission, respectively. An intense radiation mode emerges within an angular spread $\Delta\theta<3^\circ$, corresponding to a standard deviation $\Delta k = 0.3$ $\mu$m$^{-1}$ in k-space. Approximating such a partially coherent beam as a Gaussian Schell-model source \cite{friberg1982}, we can determine a spatial coherence length of 4 $\mu$m, which is close to the spatial dimension of the lasing mode (Fig.~\ref{fig:kr_images}a inset). On the contrary, no lasing action occurs for the cross-circular $\sigma^+/\sigma^-$ component, which exhibits an angularly broad intensity distribution and a parabolic $E$ vs. $k_\parallel$ dispersive spectrum. Accordingly, the radiation at the threshold has a unity circular polarization.

Next, we describe the nonlinear input-output and polarization characteristics with varying pump flux. Fig. \ref{fig:lasing}a shows the emission flux (output) vs. the pump flux (input) under a circularly ($\sigma^+$) polarized excitation. The pump flux, $P$, is the photon flux per pulse transmitted into the microcavity within a circular 10 $\mu$m diameter area. The output nonlinearly increases by one order of magnitude for an increase in the input by less that 20\% near the critical photoexcited density. The onset of such a nonlinear output for the co-circular component ($\sigma^+/\sigma^+$) is defined as the threshold $P_{th}$ (indicated by an arrow in the figure). For $P \gtrsim 1.5\,P_{th}$, the cross-circularly polarized component also lases. Under a linearly polarized pump, the laser action commences at a slightly higher pump flux ($P = 1.05 \, P_{th}$) (not shown, see Ref. \cite{hsu2013a}). This 5$\%$ threshold reduction with the optical injection of the spin-polarized carriers is small but significant compared with the $<$1$\%$ reduction predicted for an InGaAs-MQW-based VCSEL \cite{oestreich2005}. In general, such a threshold reduction is less than 5$\%$ in most locations and samples studied in this work.

The total emission under a circularly polarized pump is close to that under a linearly polarized one. The overall efficiency (the ratio of the emission flux emanating from the front surface [output] to the pump flux transmitted into the microcavity [input]) reaches a plateau of $\sim10\%$ at $P\gtrsim3\,P_{th}$. In the plateau regime, the output linearly increases with the input and resembles the characteristics of a conventional semiconductor laser. The maximal efficiency ranges from $3\%$ to $11\%$. An efficiency greater than $10\%$ can be obtained. Absorption in the nine 6-nm thick In$_{0.15}$Ga$_{0.85}$As/GaAs MQWs in the cavity is $12\%$ at $\lambda_p=785$ nm at room temperature. Therefore, an efficiency greater than $10\%$ implies that essentially all of the carriers photoexcited in the MQWs can recombine radiatively and contribute to laser action.

The spontaneous build-up of the circularly polarized radiation at a critical photoexcited density can be quantified by the $DoCP$ ($\bar{\rho}_c$) (Fig.~\ref{fig:lasing}b) deduced from the normalized Stokes vector $s = \{s_1, s_2, s_3\}$ (see Appendix: Methods). Below the threshold, the radiation is unpolarized ($\bar{\rho_c} \approx$  0). Slightly above the threshold ($P_{th} < P <1.2 \ P_{th}$), the radiation is highly circularly polarized ($\bar{\rho}_c >$ 0.95). For $P>1.5\,P_{th}$, the radiation becomes elliptically polarized with reduced $\bar{\rho_c}$ as a result of increasing radiation with an opposite helicity. When the helicity of the circularly polarized pump is switched, $\bar{\rho_c}$ changes in sign but maintains the same magnitude, i.e., the polarization state is symmetric with respect to the helicity of the pump. The pump flux-dependent $DoCP$ is quantitatively reproduced (dashed lines in Fig.~\ref{fig:lasing}b) by a rate-equation model assuming a spin- and density-dependent stimulated process, as described in the Appendix.
 
\begin{figure}[htb!]
\centering
\includegraphics[width= 0.34\textwidth]{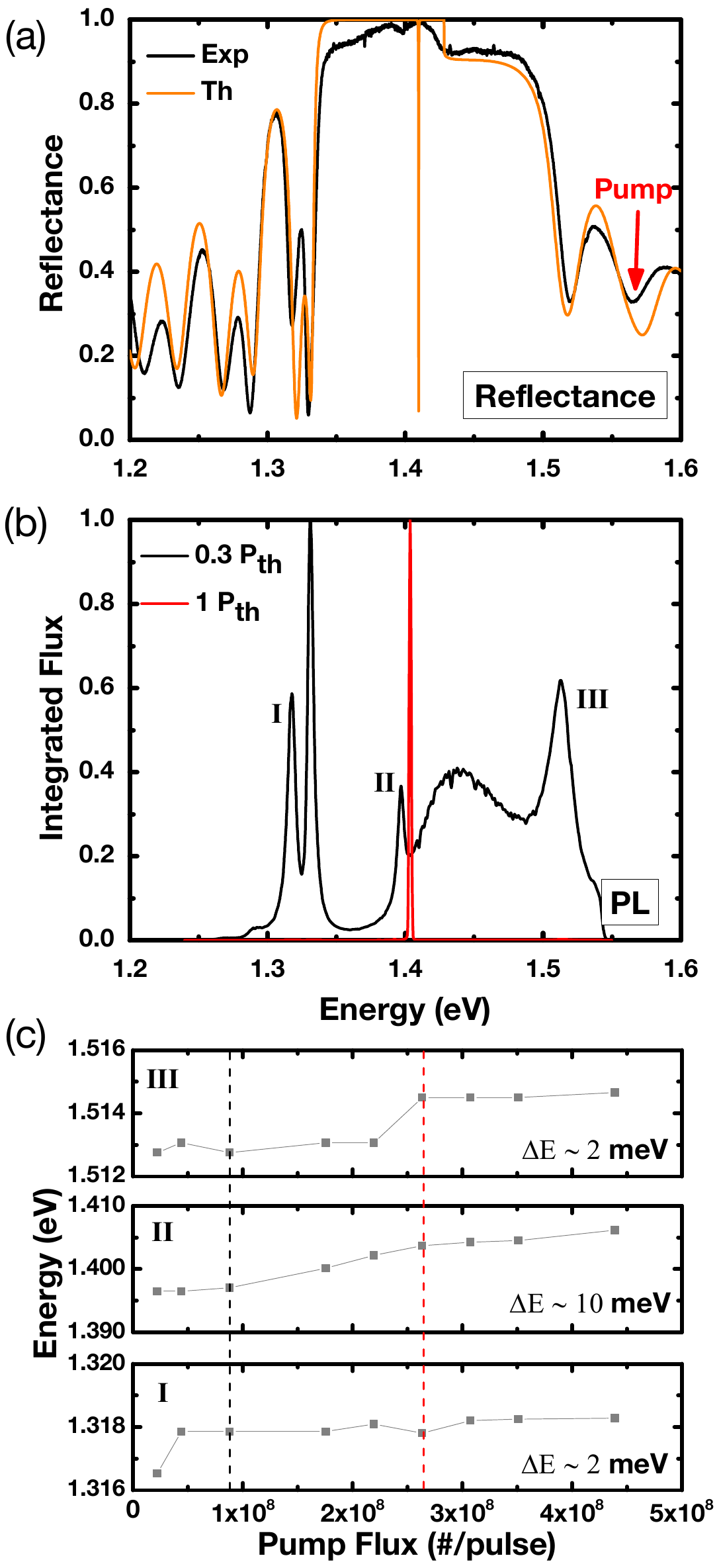}
\caption{\label{fig:ref}\textbf{Reflectance and laser spectral characteristics.} (a) Reflectance spectrum at $k_\parallel = 0$ from the front surface of a microcavity sample with 1.405 eV lasing energy at the threshold: measured (black solid line) and simulated (orange solid line). This sample is an as-grown sample not subject to a rapid thermal annealing process. The simulation is performed via a transfer matrix method, including the optical absorption in the GaAs layers, but excluding the complex dielectric constant of excitons (\emph{e-h} pairs) in MQWs. The cavity resonance ($E_c$) is about 1.41 eV. (b) Emission spectra at $k_\parallel = 0$ at at pump $P$ = 0.3 and 1.0 $P_{th}$. Above the threshold, luminescence is dominated by the lasing mode (red curve). (c) Spectral peak energy as a function of pump flux (photoexcited density) for emission peaks I, II, and III as indicated in (b). The vertical black and red dashed lines indicate $P =$ 0.3 and 1.0 $P_{th}$, respectively. Emission peak I and peak III correspond to a reflection minimum with energy below and above the reflection stop band, whereas peak II is near the cavity resonance. Peak I and peak III experience less than 2 meV spectral shifts with an increasing pump flux. By contrast, peak II blueshifts more than 10 meV for the same density range, a resulting indicating a strong optical nonlinearity induced by the coupling between the high-density \emph{e--h} plasma and the cavity light field.
}
\end{figure}

In Fig.~\ref{fig:spec}, we study the spectral characteristics. When the pump flux is increased from $P$ = 0.5 $P_{th}$ toward the threshold, luminescence blueshifts by $\approx$5 meV, whereas the linewidth $\Delta E$ decreases from about 10 meV to 0.3 meV. The linewidth $\Delta E$ [full width at half maximum (FWHM)] of the spectral distribution and the peak energy of the co- and cross-circularly polarized spectra at $k_\parallel=0$ under a circularly polarized ($\sigma^{+}$) pump are plotted in Fig.~\ref{fig:spec}b. Slightly above the threshold ($P_{th}<P<1.5\,P_{th}$), spectrally narrow ($\Delta E \approx$ 0.3 -- 1.0 meV) radiation emerges with a nonlinear growth in magnitude, whereas the peak energy remains constant. Far above the threshold ($P >1.5\,P_{th}$), the spectral linewidth increases to more than 2 meV. The overall emission energy shifts with the increasing pump flux are more than 10 meV (Fig.~\ref{fig:spec}c), which is significantly larger than the corresponding energy shift of the cavity stop band ($<$2 meV) (Appendix Fig.~\ref{fig:ref}). In addition, far above the threshold, co- and cross-circular components both lase with the rising peak energy while retaining an energy splitting of $\approx$1--2 meV (Fig.~\ref{fig:spec}d). 

\begin{figure}[htb]
\includegraphics[width=0.35\textwidth]{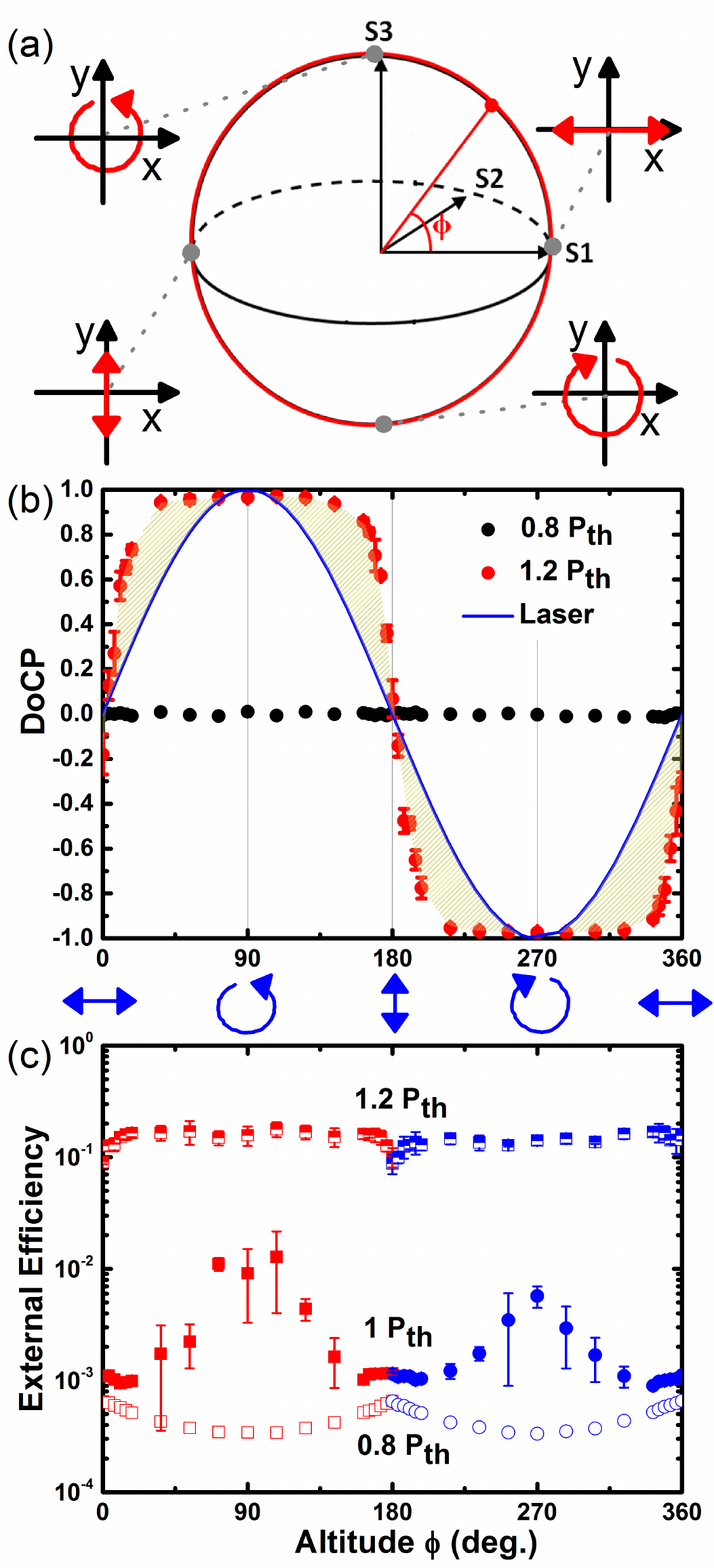}
\caption{\label{fig:spin_amp}\textbf{Spin amplification under an elliptically polarized pump.} (a) Representation of polarization states (Stokes vectors) in a Poincar\'e sphere. The pump polarization is varied along the meridian in the $s_1$-$s_3$ ($x$-$z$) plane. (b) The time-integrated $DoCP$ ($\bar{\rho}_c$) of the spin-polarized laser radiation as a function of pump $DoCP$ ($\rho_c^{p}$) (blue line) at 0.8 (black dots), 1.0 (magenta dots), and 1.2 $P_{th}$ (red dots). (c) External quantum efficiency ($\eta_{ex}$) vs. pump $DoCP$ ($\rho_c^{p}$, represented by the altitude $\phi$) at 1.2, 1.0, and 0.8 $P_{th}$. The pump flux is maintained at a constant when $\rho_c^{p}$ is varied. For a specific $\phi$, only the $\eta_{ex}$ of the majority polarized emission component is shown ($\sigma^+$ for $0^\circ \leq \phi < 180^\circ$ and $\sigma^-$ for $180^\circ \leq \phi < 360^\circ$). $\eta_{ex}$ of the minority component is not shown because of a low signal-to-noise ratio for $\rho_c^p \sim \pm1$. Error bars represent the standard deviation of $\eta_{ex}$ over five measurements.} 
\end{figure}
 
\begin{figure}[htb]
\includegraphics[width=0.36\textwidth]{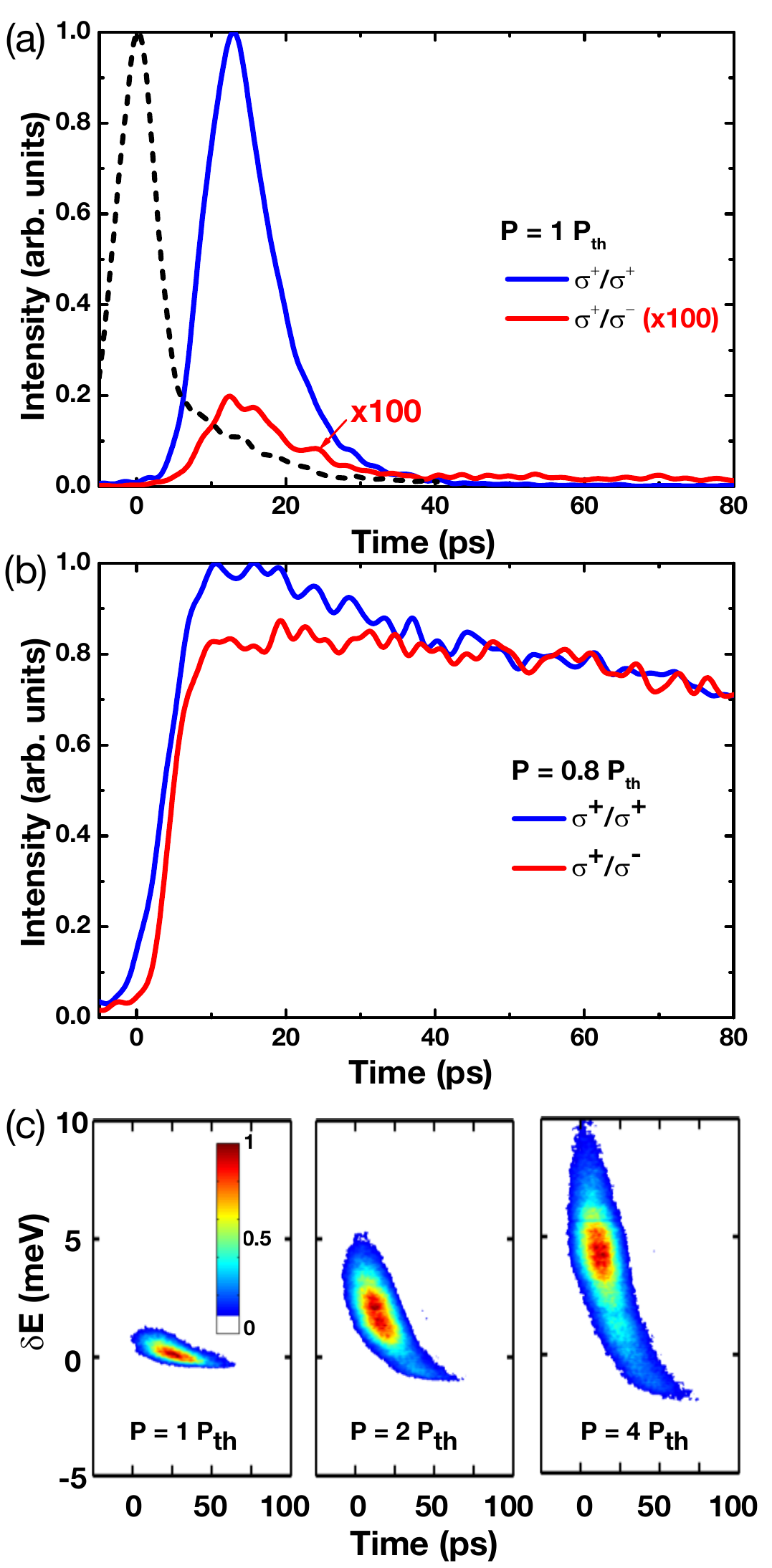} 
\caption{\label{fig:dynamics}\textbf{Dynamics and energy relaxation.} (a) Polarized time-dependent luminescence at $k_\parallel$ = 0 for $P$ = 1.0 $P_{th}$ under a circular polarized ($\sigma^+$) pump. Blue (red) curves represent the co-circular $I^+(t)$ [cross-circular $I^-(t)$] components. Note that the cross-circular component shown at $P = P_{th}$ is multiplied by a factor of 100. The time zero is determined from the instrument response (black dashed curve), which is measured via pump laser pulses reflected off the sample surface. The time traces are spectrally integrated (temporal resolution $\approx 5$ ps). (b) Same as (a) but for $P$ = 0.8 $P_{th}$. (c) Temporally and spectrally resolved streak spectral images of the co-circular component $I^+(\delta E,t)$ at $P$ = 1.0, 2.0, and 4.0 $P_{th}$. The y-axis ($\delta E$) is offset with respect to $1.408$ eV, the lasing energy at $P_{th}$. The temporal resolution is $\approx$30 ps because of the grating-induced dispersion.}
\end{figure}

Another manifestation of the spin-dependent process is highly circularly polarized lasing even under a nonresonant elliptically polarized optical excitation (Fig. \ref{fig:spin_amp}). Here, the pump flux transmitted into the microcavity sample is kept constant with varying pump circular polarization ($\rho_c^{p}$, represented by the pump Stokes vector tracing a meridian at the Poincar\'e sphere shown in Fig.~\ref{fig:spin_amp}a). When the initial spin-dependent population imbalance is controlled by variation of $\rho_c^{p}$, the $DoCP$ of the lasing radiation ($\bar{\rho}_c$) can exceed that of the pump for $1.0 \, P_{th}< P < 1.5 \,P_{th}$ (Fig. \ref{fig:spin_amp}b). Such a spin amplification arises from the ``gain" anisotropy in the presence of two threshold pump fluxes for two helicities of laser radiation, typically denoted as $JT1$ and $JT2$ in the literature on spin-controlled VCSELs \cite{gothgen2008,*zutic2011,*lee2012,*lee2014}. Highly circularly polarized lasing ($\bar{\rho}_c >$  0.8) occurs even when the $\rho_c^{p}$ is as low as 0.5 (altitude $\phi$ = 30$^\circ$). Next, we consider a polarization-dependent external efficiency ($\eta_{ex}^\pm$) (Fig.~\ref{fig:spin_amp}c), which is defined as the ratio of the polarized emission flux emanating from the front surface (output) to the pump flux transmitted into the microcavity of the same polarization (input). The external efficiency $\eta_{ex}$ of the majority polarized emission component is less than $10^{-3}$ below the threshold, and increases by two orders of magnitude at 1.2 $P_{th}$. In both cases, $\eta_{ex}$ is insensitive to $\rho_c^{p}$. At the threshold, $\eta_{ex}$ exceeds $10^{-2}$ for $\rho_c^{p} \approx$ 1 ($\phi$ = 90$^\circ$, 270$^\circ$), whereas it remains $\sim10^{-3}$ for $\rho_c^{p} \approx$ 0 ($\phi$ = 0$^\circ$, 180$^\circ$). The low $\eta_{ex}$ below the threshold is due to significant loss through nonradiative recombination, reabsorption, and emissions into other nonlasing modes. With the increasing pump flux, a stimulated process dominates over the loss and yields a dramatic increase in $\eta_{ex}$ above the threshold. A competition between loss and spin-dependent stimulation can result in the observed ``spin amplification" ($\bar{\rho_c} > \rho_c^p$) effect under elliptically polarized pumping, which is qualitatively reproduced by the aforementioned model (Appendix Fig.~\ref{fig:sim}).

To understand the mechanism of the spin-polarized laser action, studying the polarization dynamics through time-resolved polarimetry and spectroscopy is needed. Fig. \ref{fig:dynamics} shows the selected time-resolved co- and cross-circularly polarized luminescence [$I^\pm(t)$] under a $\sigma^+$ circularly polarized pump. Below the threshold, the time-dependent $\rho_c(t)$ reaches $\sim$0.1 when the luminescence reaches its peak, and then it decays with a time constant less than 10 ps, as demonstrated by the minimal transient difference between $I^+(t)$ and $I^-(t)$ at $P$ = 0.8 $P_{th}$ (Fig. \ref{fig:dynamics}b). At the threshold, the co-circular component commences the pulsed laser action within 30 ps, whereas the cross-circular component remains negligible [$I^{+}(t)/I^{-}(t)>100$] (Fig. \ref{fig:dynamics}a). Such a high $\rho_c(t) > 0.9$  is indicative of a spin-dependent stimulated process through which spin polarization is amplified \cite{bresler1987,hovel2005,*gerhardt2006,*gerhardt2011,*gerhardt2012,*hopfner2014,adams2009}.

\begin{figure}[htb!]
\centering
\includegraphics[width=0.4\textwidth]{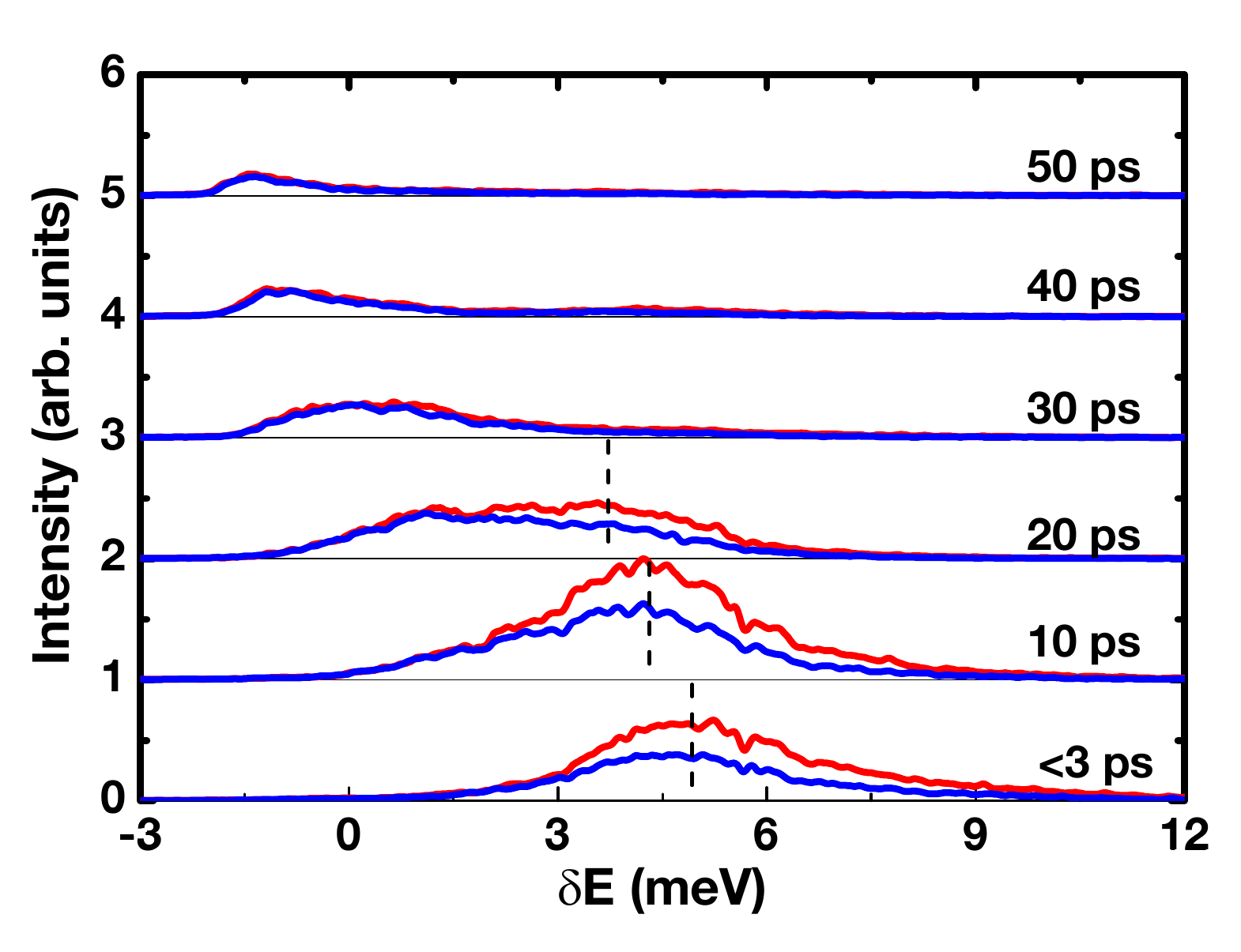}
\caption{\label{fig:trsr}\textbf{Transient lasing spectra at $P = 4 \, P_{th}$.} Cross-sectional transient spectra at specified time delays extracted from the temporally and spectrally resolved streak images of the sample. Transient spectra are averaged over 5 ps and normalized to the maximal peak intensity of the co-circular component. The spectra are equally scaled but offset vertically by 1. Red lines represent the co-circular ($\sigma^{+}/\sigma^{+}$) component, whereas blue lines represent the cross-circular ($\sigma^{+}/\sigma^{-}$) one. The energy scale is measured with respect to 1.408 eV, which is the peak lasing energy at the threshold $P_{th}$. In the initial time delay of a few ps after pulse excitation, the emission is spectrally broad and blueshifts by about 5 meV. For delays less than 20 ps, the co- and cross-circular components have the same spectral peak energy (vertical dashed lines), and this suggests that carrier interactions with the same spins and opposite spins are in a comparable magnitude. The spectral peak energy is determined by the total carrier density instead of individual spin-up or spin-down population. On the other hand, the intensity of the co-circular component is higher than that of the cross-circular component within 20 ps after pulse excitation because the ineffective spin-flipping in the reservoir results in an imbalanced spin population. The emission spectrum reaches a maximum at about 10 ps, and then gradually decreases in overall magnitude and redshifts with time. The temporal energy redshift in spectra is attributed to a descending chemical potential $\mu$ with a decreasing carrier density.}
\end{figure}

We further conduct temporally and spectrally resolved measurements, as shown in Fig. \ref{fig:dynamics}c. At the threshold, the radiation remains spectrally narrow, with a peak energy that is nearly constant with time. Above the threshold, the radiation expands spectrally when the laser action commences, and it gradually redshifts with time. In addition, polarimetric measurements reveal a circularly polarized high-energy component during the initial 10--20 ps pulsed radiation, followed by an unpolarized low-energy one for $P > 1.5 \ P_{th}$ (Appendix Fig.~\ref{fig:trsr}). The co- and cross-circular components have the same transient spectral peak energy depending on the \emph{total} carrier, and this indicates a spin-independent chemical potential $\mu$. 

\section{Discussion}

We now distinguish the present room-temperature system from exciton-polariton condensates at cryogenic temperatures \cite{deng2002,kasprzak2006,balili2007,lai2007,*utsunomiya2008,lagoudakis2008,*lagoudakis2009,*sanvitto2010,*nardin2011,*roumpos2011,bajoni2007,*bajoni2008,*assmann2011,*kammann2012,christmann2012a,schneider2013,*bhattacharya2013,deng2010,*keeling2011,*deveaud-pledran2012,*carusotto2013,*bloch2013,*byrnes2014}. A laser is typically characterized by its coherence and spectral properties, whereas a macroscopic condensate exhibits unique energy and momentum distributions. Figs.~\ref{fig:kr_images} and \ref{fig:lasing} present data on the increase of spatial coherence, nonlinear growth of macroscopic occupation in a state in energy and momentum space, and the spontaneous increase in circular polarization. We note that some exciton-polariton condensates exhibit linearly polarized radiation. This polarized radiation is a result of an energy splitting of the order of $100$ $\mu$eV between two linearly polarized modes ($\sigma^X$ and $\sigma^Y$) induced by structural disorder and strain \cite{kasprzak2006}. The angular and spectral distributions of radiation from the polariton laser studied here resemble those observed in the condensates of exciton-polaritons at cryogenic temperatures \cite{kasprzak2006,lai2007}. However, the room-temperature microcavity system presented here is a plasma laser with dynamics and spectral characteristics affected by the many-body effects at high photoexcited densities rather than the stimulation of exciton-polaritons in the strong coupling regime \cite{weisbuch1992,*houdre2002,*weisbuch2005a}.

The highly photoexcited microcavity studied in this work is one of the coupled electron-hole-photon (\emph{e--h-$\gamma$}) systems that are becoming a platform for studies on non-equilibrium collective quantum states. A many-body state near the Fermi edge in a degenerate high-density \emph{e--h} system can result in unusual optical properties. For example, Fermi-edge superfluorescence \cite{kim2013} has been observed in a degenerate \emph{e--h} system, whereas Fermi-edge polaritons \cite{gabbay2007,*smolka2014} or Mahan excitons \cite{noyes1965,*mahan1967,plochocka-polack2007} have been observed in a microcavity containing a 2D electron gas. Theoretically, it has been predicted that BCS-like states have been predicted to arise from a high-density coupled \emph{e--h-$\gamma$} system \cite{keldysh1968,*keldysh1995,*keldysh1995a,comte1982,*nozieres1982,*nozieres1985,schmitt-rink1985,*haug1984,zhu1996,*littlewood1996,*eastham2001,*keeling2005,kamide2010,*kamide2011,kamide2012,yamaguchi2012,*yamaguchi2013}.

To understand the spin-dependent polarization and spectral characteristics, we propose that when the chemical potential ($\mu$) of the degenerate \emph{e--h} plasma ($M_{eh}$) advances toward the cavity resonance ($E_c$), a fraction of the \emph{e--h} pairs near the Fermi edge can couple effectively to the cavity light field and form a coherent state ($n_0$). When the conversion from $M_{eh}$ into $n_0$ overcomes the decay of $n_0$, the occupation number in $n_0$ approaches unity. Subsequently, a stimulated process ($\propto M_{eh} n_0$) prevails and results in a nonlinear population increase in $n_0$ and in laser radiation with an increasing $M_{eh}$. 

In this microcavity, the lasing polarization is determined by $n_0$ as a result of spin-preserved stimulation. The cooling time of nonresonantly photoexcited carriers is comparable to the spin relaxation time ($\tau_s$) (Appendix Fig.~\ref{fig:dynamics}b). Consequently, under a circularly polarized $\sigma^+$ pump, a sizable spin-imbalanced population is built up when carriers are cooled down to $M_{eh}$. In the absence of stimulation (below the threshold), radiation from $n_0$ is long-lived because of the relatively long population decay time of $M_{eh}$ ($\sim100 \ \tau_s$). Therefore, the time-integrated circular polarization $\bar{\rho}_c$ is close to zero (unpolarized). In this regime, the nonradiative loss dominates because of inefficient conversion from $M_{eh}^\pm$ to radiative $n_0^\pm$. As a result, the overall radiative efficiency of the \emph{e--h} system is low. When the stimulation condition is satisfied for $M_{eh}^+$ but not $M_{eh}^-$, the conversion rate from $M_{eh}^+$ to radiative $n_0^+$ increases rapidly and yields a macroscopic population in $n_0^+$ and a high overall radiative efficiency. By contrast, the overall radiative efficiency for $n_0^-$ remains minimal. Therefore, the resulting \emph{pulsed} laser radiation is circularly polarized with near unity $\bar{\rho}_c$. Such fully circularly polarized lasing only occurs in a limited photoexcited regime, which is determined by the finite spin-imbalance between $M_{eh}^+$ and $M_{eh}^-$ when carriers are cooled down. When both $M_{eh}^+$ and $M_{eh}^-$ reach the stimulation condition at a high photoexcited density, the laser radiation displays a reduced $\bar{\rho}_c$. Temporally, the decreasing $\mu$ manifests in radiation redshifts (Fig.~\ref{fig:dynamics}c), which partially contribute to the observed linewidth broadening in the time-integrated spectra (Fig.~\ref{fig:spec}a). The temporal evolution of $\mu$ and the spin population imbalance also account for an apparent spin-dependent energy splitting of about 1--2 meV (Fig.~\ref{fig:spec}d). Initially, the lasing from $n_0^+$ at a high energy dominates because the spin-flipping from $M_{eh}^+$ to $M_{eh}^-$ is minimal. With an increasing time delay, the spin-flipping from $M_{eh}^+$ to $M_{eh}^-$ becomes significant, and the laser radiation from both $n_0^+$ and $n_0^-$ appear at a lower energy.   

In the Appendix, we provide a phenomenological rate-equation model that includes spin-dependent stimulation and loss. The model reproduces quantitatively the spin-polarized lasing as a function of photoexcited density and pump polarization. In principle, the high-density \emph{e--h}-plasma lasing described in this study can be modeled by a self-consistent numerical analysis based on Maxwell-Bloch equations beyond a phenomenological spin-flip model \cite{san-miguel1995,*martin-regalado1997,*van-exter1998} developed for conventional semiconductor lasers \cite{schmitt-rink1986a,*koch1995,*sarzala2012,*debernardi2013}; however, the strong optical nonlinearities induced by the coupling of the \emph{e--h} plasma and the cavity light field should considered. For example, an index-induced cavity resonance shift can be considered a mechanism for the observed density- and time-dependent lasing energy shift. The cavity resonance shift ($\delta E_c$) is related to the change in the refractive index ($\delta n_c$) approximately through $\delta E_c/E_c = - \delta n_c/n_c$, where $n_c$ is the effective refractive index averaged over the longitudinal cavity photon mode. Therefore, a cavity resonance shift $\delta E \sim$10 meV (Fig.~\ref{fig:spec}) requires a sizable reduction of the refractive index, i.e., $|\delta n_c/n_c| \sim$ 0.7\%. Such a significant $\delta n_c$ is probable with resonance-enhanced optical nonlinearity, which is consistent with the aforementioned framework based on the formation of cavity-induced correlated \emph{e--h} paris near the Fermi edge. Nonetheless, analyzing the polarized spectral characteristics and dynamics of lasers with highly interacting carriers in the gain media is important to further understand Coulomb many-body effects, such as screening, bandgap renormalization, and phase-space filling \cite{comte1982,*nozieres1982,*nozieres1985,schmitt-rink1985,*haug1984,schmitt-rink1985a,jahnke1995,*kira1997,*kira2006,*koch2006} in high-density coupled \emph{e--h}-photon systems \cite{keeling2005,kamide2010,*kamide2011,kamide2012,yamaguchi2012,*yamaguchi2013}. 

In the present studied spin-polarized microcavity laser, the spontaneous emission factor \cite{baba1991,*baba1992,bjork1993,*bjork1994} is $\sim10^{-2}$ and the external quantum efficiency can reach 10\%, which matches that of the carriers absorbed in the QW gain media. Moreover, the lasing threshold carrier density is nearly the same as that in conventional VCSELs even when the cavity resonance is far detuned from the bandgap of the QWs ($E_c-E_g' \sim 80$ meV). Theoretically, the threshold density is expected to increase with the increasing detuning, except when many-body interactions in the photoexcited carriers are considered \cite{kamide2012}. We attribute the high quantum efficiency at a high photoexcited density and the low threshold under a sizable detuning to the stimulation of correlated \emph{e--h} pairs formed near the Fermi edge of the high-density plasmas as a result of the coupling to the cavity light field. Moreover, the lasing energy is largely determined by the chemical potential of the plasmas rather than by bare cavity resonance; as a result, tuning the laser energy for more than 10 meV with an increasing photoexcited density is possible. Specifically, the emission energies blueshift more than $\gtrsim 10$ meV when the photoexcited density increases from $2\times 10^{12}$ cm$^{-2}$ to $\sim10^{13}$ cm$^{-2}$ per pulse. Our results suggest potential applications for wavelength tunable lasers \cite{chang-hasnain2000,coldren2004}, with polarization controlled by an external stimulus rather than being fixated to static structures. The sub-10 ps pulsed laser action commencing within 10 ps after pulse excitation suggests that a high-speed operation ($>$100 GHz) is feasible \cite{chang2013}. In addition, the studied microcavity structure with the cavity resonance tuned to the excited quantized levels of QWs can be used for spin-controlled high-speed VCSELs with electrical injection of spin-polarized carriers via ferromagnetic contacts \cite{kioseoglou2004,frougier2013,chen2014}.
 
In summary, we have described a spin-polarized laser that exhibits nonlinear energy shifts, spin-dependent energy splittings, and linewidth broadening with an increasing photoexcited density. The ultrafast (sub-10 ps) room-temperature spin-polarized lasing occurs in a highly photoexcited microcavity in which correlated \emph{e--h} pairs are formed near the Fermi edge. The spin-dependent stimulation and high optical nonlinearities arising from cavity-induced many-body states play an important role in facilitating the observed spin-polarized lasing presented in this study. Our results should stimulate activities that exploit spin and many-body effects for fundamental studies of light-matter interactions, as well as facilitate developments of spin-dependent optoelectronic devices.  

\begin{acknowledgements}
We thank Jack Bass, Mark Dykman, Brage Golding, John A. McGuire, Carlo Piermarocchi, Y. Ron Shen, and Hailin Wang for the discussions. This work was supported by the NSF DMR-0955944 and J. Cowen Endowment at Michigan State University.
\end{acknowledgements}

\begin{figure*}[htb!]
\centering
\includegraphics[width= 0.9 \textwidth]{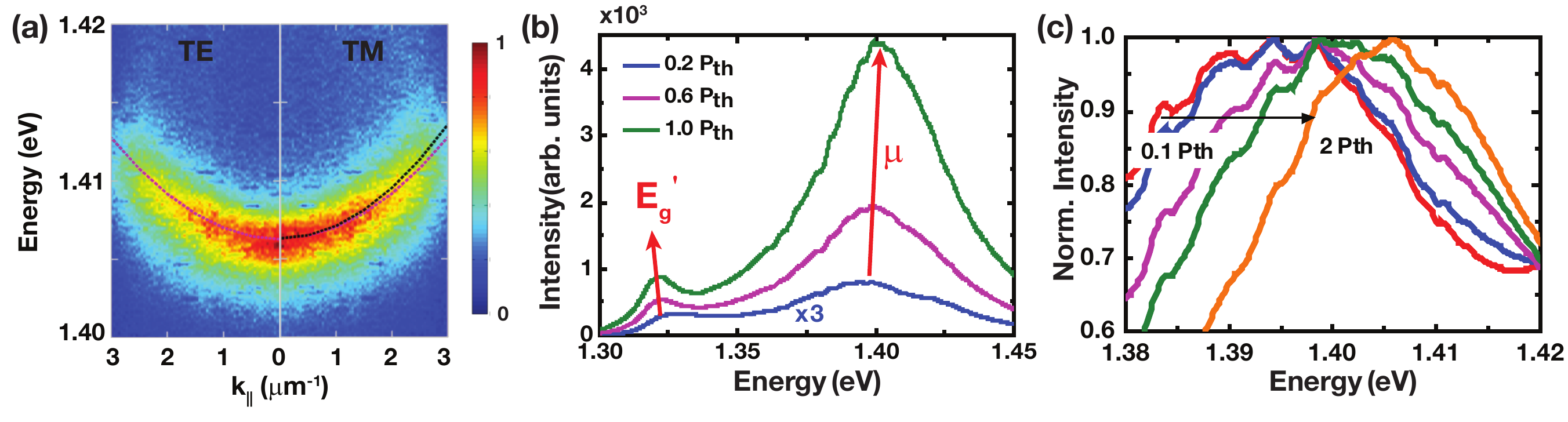}
\caption{\label{fig:sample}\textbf{Microcavity sample characterization.} (a) $E$ vs. $k_\parallel$ dispersions for the TE (left) and TM (right) modes under a circularly polarized  pump at 1.579 eV and $P$ = 0.6 $P_{th}$. The photoexcited density at $P_{th}$ is $\sim2-3 \times 10^{12}$ cm$^{-2}$ per quantum well \emph{per pulse}. The simulated TE and TM dispersions are shown as magenta and black curves, respectively. The TE-TM energy splitting is less than 50 $\mu$eV at $k_\parallel = 0$, allowing the effective control of lasing polarization by optical pumping. (b) To determine the density-dependent spectral characteristics of PL in the InGaAs/GaAs MQWs, we measure the time-integrated and time-resolved PL in the sample, with the top DBR mirror layers removed, by selective wet etching \cite{desalvo1992}. PL spectra at $P =$ 0.2 (blue), 0.6 (magenta), and 1 $P_{th}$ (green) in the absence of the top DBR mirrors. Here, PL is attributed to the spontaneous radiative recombination of photoexcited carriers in the InGaAs/GaAs MQWs. The dual PL spectral peaks are attributed to the first and second quantized energy levels in the MQWs, respectively. The QW bandgap ($E_g'$) corresponds to the ground state, the transition between the first quantized energy levels of the electron and the heavy-hole ($e1hh1$). $E_g'$ can decrease with the increasing photoexcited density (band gap renormalization). $\mu$ can be deduced from PL from the excited state (second quantized levels, $e2hh2$ transition). With an increasing density, ${E_g'}$ redshifts slightly because of band gap renormalization, whereas $\mu$ blueshifts considerably ($\sim$ 10 meV) as a result of phase space filling (Pauli blocking) at a high photoexcited density ($\gtrsim 10^{12}$ cm$^{-2}$ per QW). (c) Normalized PL spectra near 1.40 eV, which displays a significant spectral blueshift of 15--20 meV with the increasing photoexcited density (0.1 to 2 $P_{th}$). 
}
\end{figure*}

\appendix
\section{Appendix: Methods}\label{sec:methods}
\textbf{Sample fabrication.} We grow the microcavity on a semi-insulating (100)-GaAs substrate by using a molecular beam epitaxy method. The top (bottom) DBR consists of 17 (20) pairs of GaAs(61-nm)/AlAs (78-nm) $\lambda/4$ layers. The central cavity layer consists of three stacks of three In$_{0.15}$Ga$_{0.85}$As/GaAs (6-nm/12-nm) quantum wells each, positioned at the anti-nodes of the cavity light field. The structure is entirely undoped and contains a $\lambda$ GaAs cavity sandwiched by DBRs; the result is a bare cavity resonance $E_c \approx$ 1.41 eV ($\lambda_c$ = 880 nm) at room temperature (Fig.~\ref{fig:ref}). The QW  bandgap ($E_g' \sim$1.33 eV) is tuned through a rapid thermal annealing process (at 1010 $^\circ$C -- 1090$^\circ$C for 5-10 s), in which the InGaAs QW bandgap blueshifts because of the diffusion of gallium ions into the MQW layers. The cavity quality factor Q is about 4000--7000, which corresponds to a cavity photon lifetime $\sim$2 ps.
 
\textbf{Thermal management.} At a high pump flux, the steady-state incident power transmitted to the sample at the 76 MHz repetition rate of the laser will exceed 50 mW and result in significant thermal heating and carrier diffusion. The techniques we use to control the thermal heating and diffusion of the photoexcited carriers are (1) temporally modulating the pump laser intensity to suppress the thermal carrier heating and (2) spatially shaping the pump beam profile to enable lasing in a single transverse mode. Steady-state thermal heating can inhibit laser action and lead to spectrally broad redshifted luminescence. To suppress steady-state thermal heating, we temporally modulate the 2-ps 76 MHz pump laser pulse train with a duty cycle (on/off ratio) $<$ 0.5 $\%$ by using a double-pass acousto-optic modulator system \cite{donley2005}. The time-averaged power is limited to below 0.2 mW for all experiments. Multiple transverse modes can simultaneously lase because of the diffusion of photoexcited carriers and crystalline disorder, which lead to instability and complex lasing characteristics. To control carrier diffusion, we holographically generate a flat-top pump beam profile (area $\approx$ 300 $\mu$m$^2$) at the sample surface with a spatial beam shaper consisting of a two-dimensional (2D) liquid-crystal spatial light modulator (SLM) \cite{romero1996}. 

\textbf{Optical excitation.} The front surface of the microcavity is positioned at the focal plane of a high-numerical-aperture microscopy objective (N.A. = 0.42, $50\times$, effective focal length 4 mm). A 3$\times$ telescope, a Faraday rotator, a polarizing beam splitter, and the objective form a reflected Fourier transform imaging system. The light fields at the SLM and sample surface form a Fourier transform pair. The 2D SLM ($1920 \times 1080$ pixels, pixel pitch = 8 $\mu$m) enables us to generate arbitrary pump geometries with a $\approx$2 $\mu$m spatial resolution at the sample surface using computer-generated phase patterns. The pump flux can be varied by more than two orders of magnitude with the use of a liquid-crystal attenuator.
 
\textbf{Imaging spectroscopy.} We measure the angular, spectral, and temporal properties of luminescence in the reflection geometry. In a planar microcavity, carriers coupled to the cavity light field are characterized by an in-plane wavenumber $k_\parallel = k \, \sin (\theta)$ because of the 2D confinement of both photons and carriers. The leakage photons can thus be used to directly measure the angular distribution of optically active carriers. Angle-resolved luminescence images and spectra are measured through a Fourier transform optical system. A removable $f = 200$ mm lens enables the projection of either k-space or r-space luminescence on to the entrance plane or slit of the spectrometer. Luminescence is collected through the objective, separated from the reflected specular and scattered pump laser light with a notch filter, and then directed to an imaging spectrometer. A single circular transverse lasing mode with a spatial mode diameter $\approx$8 $\mu$m is isolated for measurements through a pinhole positioned at the conjugate image plane of the microcavity sample surface. The spectral resolution is $\approx$0.1 nm (150 $\mu$eV), which is determined by the dispersion of the grating (1200 grooves/mm) and the entrance slit width (100--200 $\mu$m). The spatial (angular) resolution is $\approx$ 0.3 $\mu$m (6 mrad) per CCD pixel.

\textbf{Polarization control and notation.} The polarization state of the pump (luminescence) is controlled (analyzed) by a combination of liquid crystal devices, such as variable retarders and polarization rotators, and Glan-Taylor/Glan-Thomson polarizers without mechanical moving parts. A polarization compensator (Berek's variable wave plates) is used to compensate for the phase retardance induced by the reflection from the miniature gold mirror surface. The circularly polarized pump or luminescence with angular momentum $+\hbar$ ($-\hbar$) along the pump laser wavevector $\hat{k} \parallel \hat{z}$ is defined as $\sigma^{+}$ ($\sigma^{-}$). Linearly polarized light with horizontal (vertical) polarization is defined as $\sigma^X$ ($\sigma^Y$). The polarization state is characterized by the Stokes vector $\{S_0, S_1, S_2, S_3\}$. $S_0$ is the flux and is determined as $S_0 = I^{+}+I^{-} = I^{X}+I^{Y} = I^{45^\circ}+I^{135^\circ}$. The Stokes vector can be normalized by its flux $S_0$ to the Stokes three-vector $s = \{s_1, s_2, s_3\}$. $s_1 = (I^{X}-I^{Y})/(I^{X}+I^{Y})$, $s_2 = (I^{45^\circ}-I^{135^\circ})/(I^{45^\circ}+I^{135^\circ})$, and $s_3 = (I^{+}-I^{-})/(I^{+}+I^{-})$. $I^{+}$, $I^{-}$, $I^{X}$, $I^{Y}$, $I^{45^\circ}$, as well as $I^{135^\circ}$ are measured time-integrated or temporal intensities of the circular or linear polarized components. We represent the polarization state with the following three quantities: the degree of circular polarization ($DoCP \equiv \rho_c = s_3$), the degree of linear polarization ($DoLP \equiv \rho_l =\sqrt{s_1^2+s_2^2}$), and the degree of polarization ($DoP \equiv \rho = \sqrt{s_1^2+s_2^2+s_3^2}$). The accuracy of the measured polarization state is $\approx$1\%--2\%.

\section{Appendix: Rate-equation Model}\label{sec:modeling}
\begin{figure*}[htb!]
\centering
\includegraphics[width=0.9\textwidth]{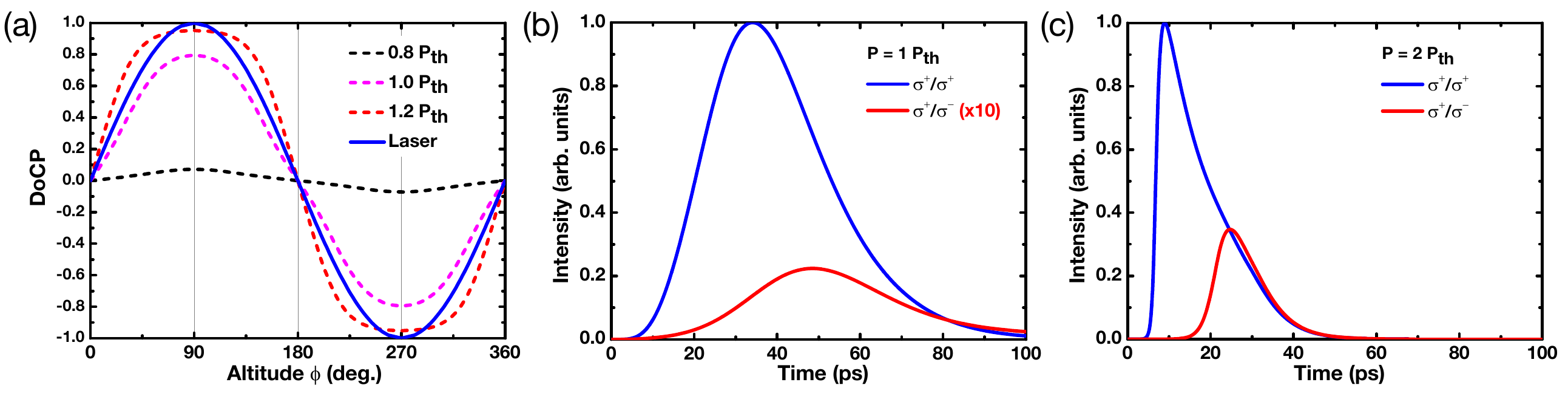}
\caption{\label{fig:sim}\textbf{Spin-dependent Stimulation.} (a) Calculated stationary $DoCP$ ($\bar{\rho_c}$) with varying pump polarization $\rho_c^{p}$ (represented by the altitude) for $P$ = 0.8, 1.0 and 1.2 $P_{th}$. (b-c) Calculated polarized time-dependent radiation intensity for co-circular $\sigma^+/\sigma^+$ [$I^+(t)$] and cross-circular  $\sigma^+/\sigma^+$ [$I^-(t)$] components under a $\sigma^+$ pump for $P$ = 1.0 and 2.0 $P_{th}$. The theoretical $P_{th}$ is set to the pump flux when the stationary $\bar{\rho}_c$ = 0.5 under a fully circularly polarized pump ($\rho_c^{p}$ = 1). }
\end{figure*}

To model the spin-controlled lasing processes, we use a set of rate equations with \emph{spin-dependent} stimulated processes. We consider two states populated with spin-polarized electron-hole (\emph{e--h}) pairs: a nonradiative \emph{e--h} plasma reservoir ($M_{eh}^\pm$) and radiative correlated \emph{e--h} pairs at $k_\parallel \approx 0$ ($n_0^\pm$). The \emph{e--h} pairs nonresonantly photoexcited by a 2 ps pulsed pump laser at 1.58 eV relax rapidly ($<$5 ps) to the reservoir via, for example, a spin-preserved scattering process with LO optical phonons. Therefore, we assume that spin-polarized \emph{e--h} pairs are optically injected into the reservoir at a generation rate of $G_p P^\pm$, where $P^\pm$ is the pump laser flux of helicity $\pm$. The spin flipping time of the \emph{e--h} pairs, $1/W_{sf}$, is less than 10 ps, as demonstrated by the polarized time-dependent PL below the threshold (Fig.~\ref{fig:dynamics}) as well as in a sample without the top DBR mirror layers (Fig.~\ref{fig:sample}). The \emph{e--h} carriers in the reservoir can also dissipate through reabsorption and nonradiative recombination ($\Gamma_{loss}$). We further assume that a fraction of \emph{e--h} pairs near the Fermi-edge couple effectively to the cavity light field and form correlated \emph{e--h} pairs ($N_{eh}^\pm = \beta M_{eh}^\pm$). The conversion of carriers from $N_{eh}$ to the radiative $n_0$ state is enabled by the following processes: (a) $W_k$, a spontaneous conversion from $N_{eh}$, and (b) $W_{ss}$, a spin-dependent stimulated scattering from $N_{eh}$. The $n_0$ state contributes to the leakage photons measured experimentally at a rate associated with the cavity photon decay rate $\Gamma_c$. The lasing dynamics can then be described by the following set of coupled rate equations:
\footnotesize
\begin{align} \nonumber
\frac{d \,M_{eh}^\pm(t)}{dt} &= G_p(t) P^\pm \mp W_{sf} \left[M_{eh}^+(t) - M_{eh}^-(t)\right] - \Gamma_{loss} M_{eh}^\pm(t)\\ \nonumber
&- W_{ss}(n_0^\pm) \, N_{eh}^\pm(t) \, n_0^\pm(t) - W_k \, N_{eh}^\pm(t) \ ,\\ \nonumber
\frac{d \,n_0^\pm(t)}{dt} &=  W_{ss}(n_0^\pm) \, N_{eh}^\pm(t) \, n_0^\pm(t) + W_k \, N_{eh}^\pm(t) - \Gamma_c \ n_0^\pm(t).
\end{align}
\normalsize

The generation rate of carriers, $G_p(t)$, from a 2 ps pulsed laser pump is represented by a Gaussian distribution with a standard deviation $\sigma$ = 2 ps. The spin-dependent stimulation rate $W_{ss}=W_0 \times (1 - n_0^\pm/n_{sat})$ is phenomenologically set to decrease with density, where the saturation density $n_{sat}$ is obtained by fitting of the pump flux-dependent, stationary $DoCP$ ($\bar{\rho_c}$). 

The fittings parameters are as follows: $W_0 = 1/10$ [ps$^{-1}$], $W_{sf} = 1/10$ [ps$^{-1}$], $W_k = 10^{-4}$ [ps$^{-1}$], $\Gamma_{loss} = 1/1000$ [ps$^{-1}$], $\Gamma_c = 1$ [ps$^{-1}$], $\beta$ = 0.015, and $n_{sat}$ = 200. Given a spatial mode area of 10--20 $\mu$m$^2$ for the $n_0$ state, the calculated threshold density is about $5-10 \times 10^{5}$ $\mu$m$^{-2}$, consistent with experimentally measured carrier density per quantum well at the threshold. This model reproduces the polarized laser output fluxes ($I^\pm$) and $DoCP$ ($\bar{\rho_c}$) as a function of pump flux $P$ (Fig.~\ref{fig:lasing}) and as a function of pump polarization $\rho_c^{p}$ (Fig.~\ref{fig:spin_amp} and Fig.~\ref{fig:sim}a). The polarized time-dependent PL is also reproduced qualitatively (Fig.~\ref{fig:dynamics} and Fig.~\ref{fig:sim}b--c).
 
We note a few limitations of this simplified model: (1) It does not consider energy relaxation and spatial carrier diffusion properly. As a result, the spectral-dependent stationary $\bar{\rho}_c(E)$ and time-resolved $\rho_c(E, t)$ cannot be reproduced. (2) The model does not consider optical selection rules under a nonresonant optical pumping and neglects the fact that hole spin relaxation time is sub-10 ps \cite{amand1994,*hilton2002}. In InGaAs/GaAs quantum wells, the initial $\rho_c$ of the band-edge PL is limited to 0.25 because of $\sim$ps hole spin relaxation for a nonresonant optical excitation with more than 150 meV excess energy in our experiments; the model uses a $\rho_c(t=0)$ = 1. Therefore, the model produces a stationary $\bar{\rho}_c$ higher than the experimental values below the threshold. Moreover, the calculated $\bar{\rho}_c$ across the threshold is smooth, which is in contrast to the ``step-like" increase with the increasing pump flux in the experiments. (3) We model the saturation effect by limiting the stimulated scattering rate $W_{ss}$ above a saturation density ($n_{sat}$). The $W_{ss}(n_0)$ can exhibit a more complex density dependence when energy relaxation and spectral broadening of the lasing mode are considered. Despite of aforementioned limitations, this simplified model can qualitatively simulate the main experimental results including polarization amplification under elliptically polarized pumping and spin-polarized pulsed lasing dynamics.
\bibliography{/Users/cwlai/Dropbox/Bib/lai_lib}

\end{document}